\documentclass{webofc}
\usepackage[varg]{txfonts}   
\begin{document}
\title{Recent selected theory developments for NICA}
%
%

\author{\lastname{David Blaschke}\inst{1,2,3}\fnsep\thanks{\email{david.blaschke@ift.uni.wroc.pl}} 
}

\institute{Joint Institute for Nuclear Research,  Dubna, Russia
\and
           University of Wroclaw,  Wroclaw, Poland 
\and
          National Research Nuclear University (MEPhI), Moscow, Russia
          }

\abstract{%
In this contribution I present a few selected topics of recent theoretical developments of relevance for 
the NICA facility under construction.
In a first part, I discuss new aspects of the QCD phase diagram like the possible existence of a critical endpoint of first-order phase transitions from the perspective of generalizations of the 3-flavor PNJL model including the conjecture of a universal pressure for the onset of deconfinement  in heavy-ion collisions and astrophysics.
A second part is devoted to first results of the newly constructed event simulator (THESEUS) which is based on the particlization of the former three-fluid hydrodynamics code by Ivanov, Russkikh and Toneev that allows also to study the role of hadronic final state interactions. Possible signals of the mixed phase accessible at NICA are considered. 
In particular, the robustness of the baryon-stopping signal of deconfinement and the occurrence of antiflow for protons have been investigated for Au+Au collisions in the range of the NICA-MPD energy scan for 
$\sqrt{s} \sim 6 \dots 8$ GeV. 
This signal is reflected also in the flow pattern of light nuclear clusters, in particular deuterons. 
The sharp peak for the $K^+/\pi^+$ ratio at $\sqrt{s} \sim 8$ GeV (the "horn" effect) is not obtained in the present version of THESEUS and calls for improvement of the equation of state input.   
	I report the recent progress in developing a generalized Beth-Uhlenbeck approach to a unified description of quark-hadron matter which includes now strangeness and reveals a new mechanism for explaining the $K^+/\pi^+$ ratio due to the pronounced occurrence of an anomalous mode in the $K^+$ at finite baryochemical potentials.
}
\maketitle
\section{Introduction}
\label{intro}
The White Paper on "Exploring Strongly Interacting Matter at High Densities" at the Nuclotron-based Ion Collider fAcility (NICA) has recently appeared as a Topical Issue in the European Physical Journal A \cite{1}. 
It comprises 56 articles related to the most prospective physics aspects to be explored at this upcoming facility. 
I shall present a few selected topics of recent theoretical developments for NICA in which I have been involved.

	First I consider key aspects of the QCD phase diagram like the possible existence of a critical endpoint (CEP) of first-order phase transitions that has been elaborated in \cite{7} for nonlocal 
Polyakov-loop extended Nambu--Jona-Lasinio (PNJL) models constrained by lattice QCD with special emphasis on the role of a repulsive vector meanfield.
	The CEP is also discussed from the perspective of generalizations of the color superconducting 
3-flavor PNJL model \cite{8} including the conjecture of a universal pressure for the onset of deconfinement  in heavy-ion collisions and astrophysics \cite{9}.
	In this context I find it worthwhile to quote results of Ref.~\cite{Fischer:2016ojn} where consequences for the QCD phase diagram have been demonstrated of enforcing a coincidence between chiral symmetry restoration and deconfinement with a correspondingly adjusted bag pressure on top of a model with dynamical chiral symmetry restoration.
These considerations of the QCD phase diagram in different dynamical models are important for the preparation of the NICA experiments as they can give theoretical estimates on the accessibility of certain QCD phases and what effects one may hope to detect in the given energy range.
  
Sect.~\ref{sec:3FH} is devoted to the new event simulation program THESEUS.	
Within a collaboration involving a team of 9 scientists from Germany, France, Italy, Poland, Russia, Ukraine and the United States we have succeeded to create a startup version for a new {\bf T}hree-fluid 
{\bf H}ydrodynamics-based {\bf E}vent {\bf S}imulator {\bf E}xtended by {\bf U}rQMD final {\bf S}tate interactions (THESEUS) and reported about its performance in \cite{2}. 
In particular, the robustness of the baryon-stopping signal of deconfinement 
\cite{Ivanov:2015vna,Ivanov:2016xev}
and the occurrence of antiflow for protons for Au+Au collisions in the range of the NICA-MPD energy scan for $\sqrt{s} \sim 6 \dots 8$ GeV \cite{2} have been reconsidered in THESEUS.
For the first time it was possible to investigate the effect of hadronic rescattering on the results of the three-fluid hydrodynamics simulations. 
To this end the ultrarelativistic quantum molecular dynamics (UrQMD) program has been implemented. 
The antiflow signal for a first order phase transition is reflected also in the flow pattern of light nuclear clusters, in particular for deuterons \cite{4}. THESEUS and its underlying three-fluid hydrodynamics model shall be developed further, in particular to be able to improve the description of the  hadronization process and capture subtle effects like the "horn" effect for the $K^+/\pi^+$ ratio.
A main goal of ongoing studies is to improve the equation of state (EoS) input.   
	In this respect recent progress in developing a generalized Beth-Uhlenbeck (GBU) approach 
\cite{Schmidt:1990,Hufner:1994ma,Blaschke:2013zaa}	
to a unified description of quark-hadron matter \cite{6} which includes now strangeness \cite{Yamazaki:2013yua,Dubinin:2016wvt} (see also \cite{Torres-Rincon:2016ahl}) has shown remarkable results.
It has revealed a new mechanism for explaining the $K^+/\pi^+$ ratio due to the pronounced occurrence of  an anomalous mode in the $K^+$ channel at finite baryochemical potentials \cite{Dubinin:2016wvt}. 
The GBU approach on the basis of a PNJL model utilizes a generalization of the $\Phi-$ derivable approach that accounts for a spectrum of hadron resonances as bound states of quarks which get dissolved into the quark-gluon plasma continuum by the Mott effect encoded in medium-dependent hadronic phase shifts. It  can describe the lattice QCD thermodynamics results on the temperature axis of the QCD phase diagram \cite{6} and should therefore next be extended to finite baryochemical potentials.  

\section{The phase diagram as probed at NICA}
\label{sec-1}
This section selects  a few aspects out of the rich thematics of the QCD phase diagram that have recently been discussed in the context of the upcoming NICA experiments. 
For a general introduction to the QCD phase diagram one may consult the review by Fukushima and Hatsuda \cite{Fukushima:2010bq} and the more recent one by Fukushima and Sasaki
\cite{Fukushima:2013rx}.

Interesting aspects that have been spared out here concern inhomogeneous condensates \cite{Buballa:2014tba} for which instead of a CEP a Lifshitz point occurs in  the phase diagram. 
Another point of interest included in the NICA White Paper is the behavior of scalar mesons in the phase diagram  \cite{Costa:2016wkj}. Their properties may trace the chiral restoration transition and provide a potentially observable signal in the two-photon channel \cite{Volkov:1997dx} if the background can be sufficiently suppressed.
Promising progress has also been made by the Functional Renormalization Group approach that provides spectral functions for the quark-meson model phase diagram \cite{Tripolt:2013jra}. In near future one can expect interesting predictions also for the part of the phase diagram probed at NICA providing the approach can be generalized to include baryons.

\subsection{CEP in a nonlocal PNJL model constrained by lattice QCD}
\label{sec-2}

Dynamical models for quark hadron matter that have the potential to provide more insights to the microphysics of the hadron-to-quark matter phase transition are based on the hierarchy of the QCD Dyson Schwinger equations where hadrons appear as bound states of quarks due to their interactions via confining forces mediated by nonperturbative gluon propagators \cite{Gao:2015kea,Roberts:2015lja,Cloet:2013jya}. The key quantities of this approach are the dynamical mass function and the wave function renormalization of the quark propagator that are described in excellent agreement with lattice QCD data
\cite{Bowman:2005vx,Parappilly:2005ei}.

\begin{figure}[!t]
\centering
\includegraphics[width=0.32\linewidth]{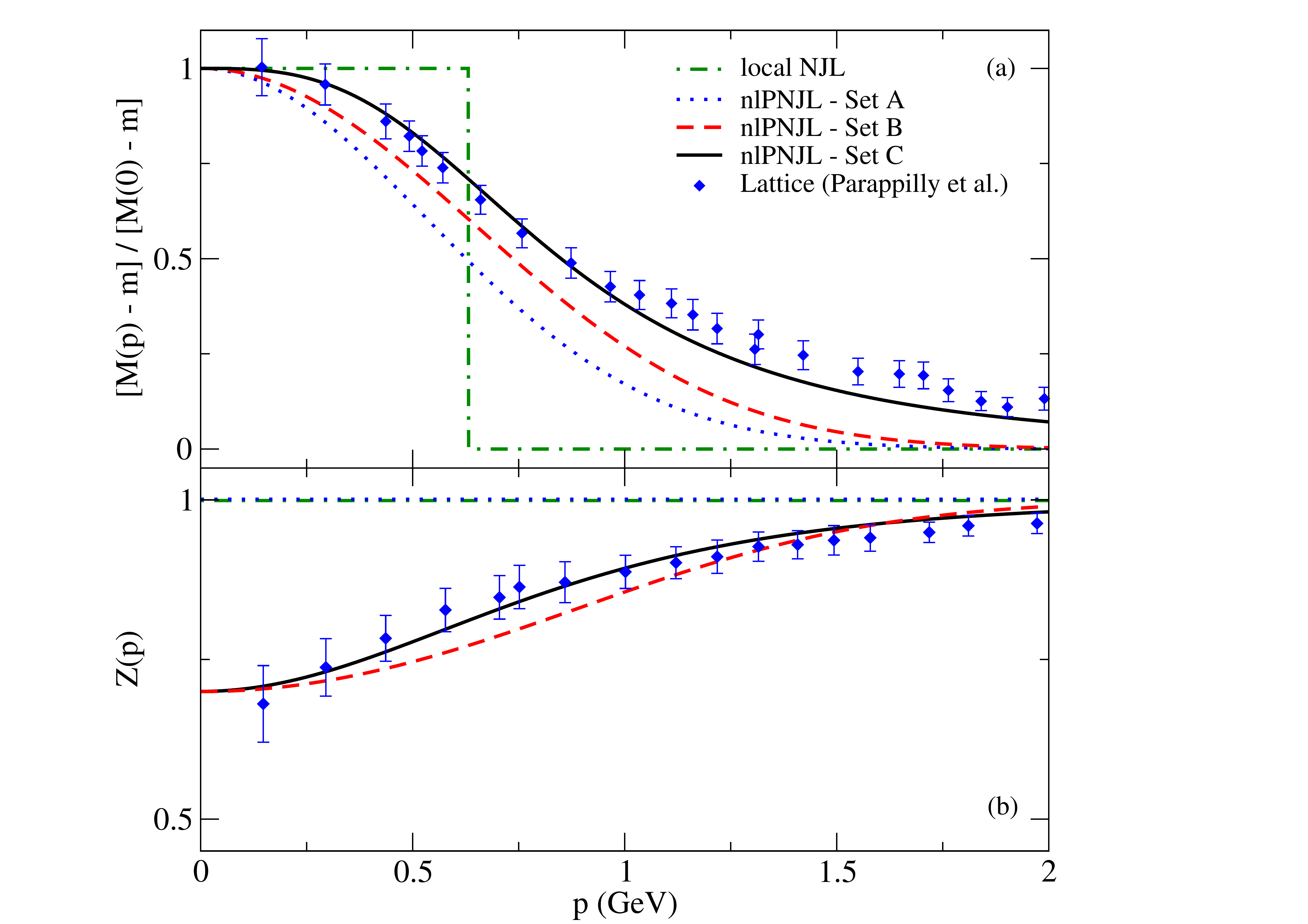}
\includegraphics[width=0.3\linewidth]{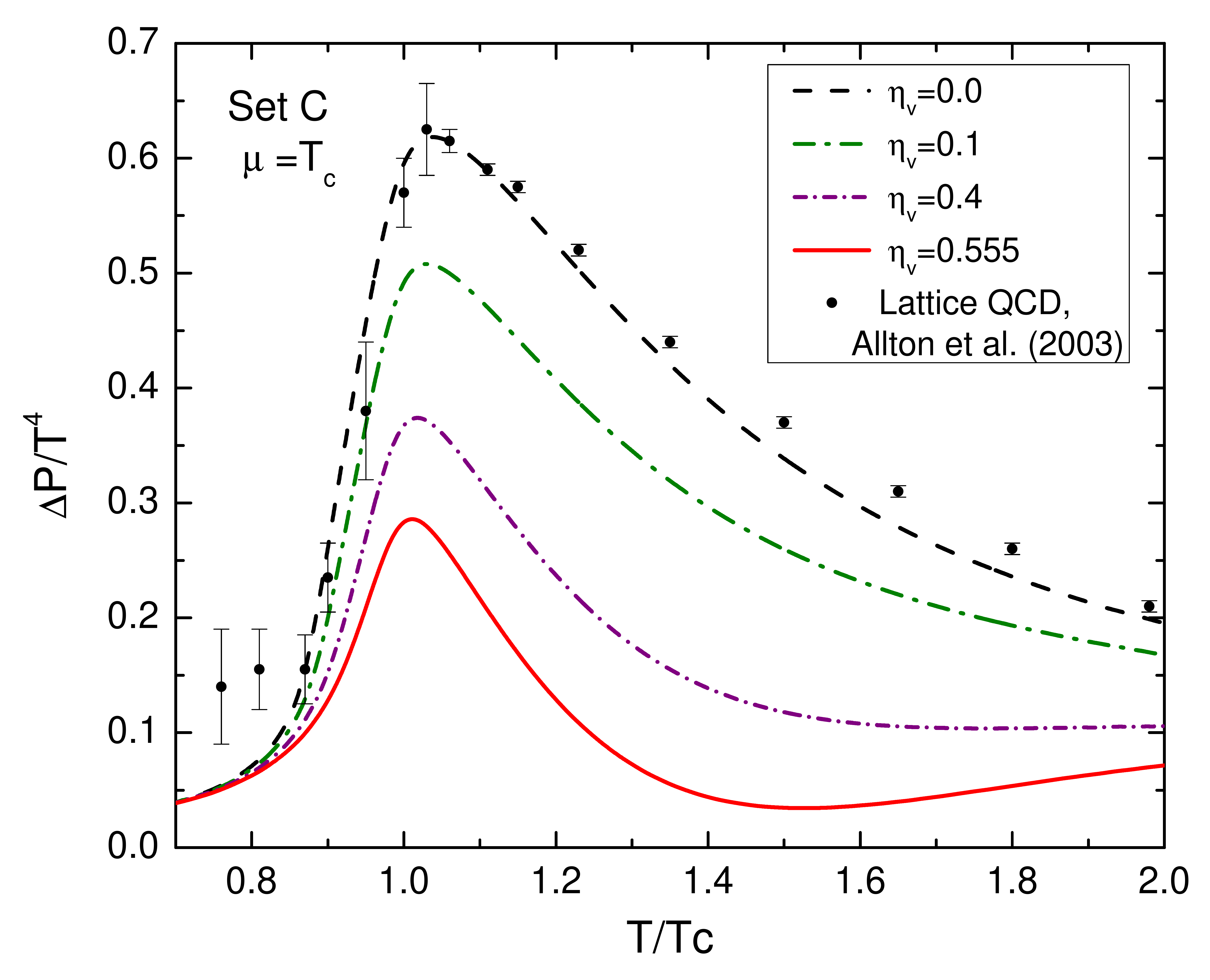}
\includegraphics[width=0.35\linewidth]{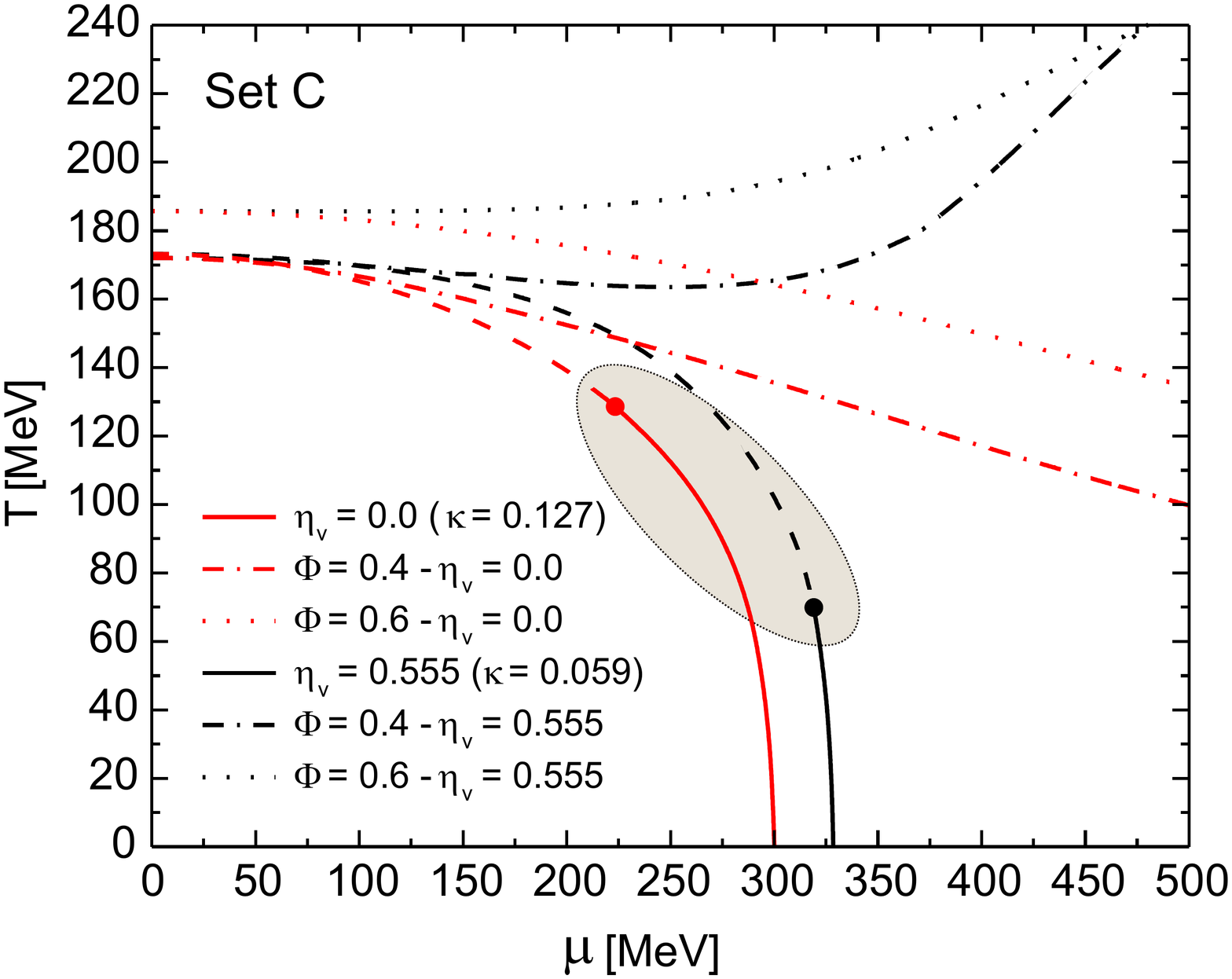}
\caption{Left panel: 
Normalized dynamical masses for the different form factors under study (NJL, Set A - C) and wave
function renormalization for Set B and Set C, fitted to lattice QCD data \cite{Parappilly:2005ei}.
Middle panel:
Comparison between Set C  results for different
values of the vector coupling parameter $\eta_V$ and lattice QCD data \cite{Allton:2003vx}.
Right panel:
Phase diagrams with (pseudo)critical temperatures $T_c(\mu)$ and critical points for nonlocal rank-2 PNJL model (Set C).
Dashed (full) lines correspond to crossover (first order) transitions. The corresponding dash-dotted and
dotted lines represent the deconfinement transition range, i.e. $\Phi=0.4$ and $\Phi=0.6$, respectively.
The highlighted region denotes the CEP position favored by the present
study.
The figures 
are from Ref.~\cite{7}.
}
\label{fig-1}       
\end{figure}

A similar quality of the description of the quark propagator is achieved within the nonlocal chiral quark models that are defined by a covariant form factor which is calibrated with the same lattice QCD data,
see the left panel of Fig.~\ref{fig-1}.
For their finite temperature extension, the coupling to the Polyakov loop results in predictions of the pseudocritical temperature that are in excellent agreement with the lattice QCD data
\cite{Horvatic:2010md,Radzhabov:2010dd}. 
The extension of these models to the whole $T-\mu$ plane has been accomplished in 
Ref.~\cite{Contrera:2012wj} with particular emphasis on the role that interactions in the Dirac vector channel play for the position and the very existence of the CEP. 
This discussion has been extended in \cite{7} where the possibility of a very small or vanishing vector coupling at low chemical potentials as obtained on the basis of lattice QCD results for the imaginary chemical potential technique has been suggested as a possible solution of the problem to describe the lattice data for the temperature dependence of the pressure difference 
$\Delta P=P(T,\mu)-P(T,0)$ at finite $\mu=T_c$ \cite{Allton:2003vx}, see the middle panel of Fig.~\ref{fig-1}. 
The favorable model is Set C which in this case predicts a critical endpoint  at $T_{\rm CEP}\sim 130$ MeV and $\mu_{\rm CEP}=225$ MeV. 
This corresponds to chemical freeze-out parameters for $\sqrt{s_{NN}}\sim 6$ GeV, i.e. below the range of the RHIC beam energy scan of the STAR experiment, but well in reach for the planned NICA energy scan.
    
\begin{figure*}[!t]
\centering
\sidecaption
\includegraphics[width=0.65\linewidth]{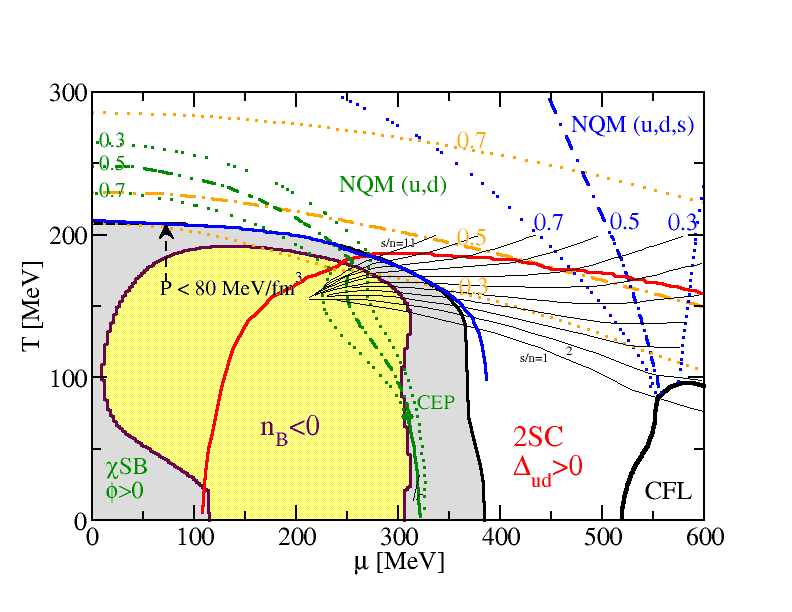}
\caption{Phase diagram of the three-flavor color superconducting PNJL model for symmetric quark matter
from \cite{8}.  For details see text in Subsect.~\ref{ssec:ayriyan}.
}
\label{fig-2}       
\end{figure*}

\subsection{Universal transition pressure in a 3-flavor  color superconducting PNJL model}
\label{ssec:ayriyan}

For the first time, the phase diagram of a 3-flavor color superconducting PNJL model has been obtained in \cite{8},  the results are shown in Fig.~\ref{fig-2}.
The entropy per baryon isolines indicate paths for the dynamical evolution of a fireball at NICA/FAIR energies where upmost line ($s/n=11$) corresponds to $\sqrt{s_{NN}}\sim 4.5$ GeV ($E_{\rm lab}\sim 9$ A GeV) \cite{Ivanov:2016hes}.
The yellow region ($n_B<0$) denotes an instability of the homogeneous meanfield approximation to quark matter which is cured when a phase transition to hadronic matter is invoked with a universal transition pressure of 80 MeV/fm$^3$ \cite{Petran:2013qla} delimiting the grey region. 
In the range of temperatures accessible in the energy scan programs of heavy-ion collisions for 
$\sqrt{s_{NN}}=4 - 200$ GeV this coincides with the phase equilibrium between the PNJL model and the hadron resonance gas (solid blue line). The solid red line corresponds to the critical temperature for 2SC superconductivity in quark matter. The green (blue) lines correspond to equal light (strange) quark masses  reduced relative to the vacuum value by a factor labelling these lines. Along the orange lines the traced Polyakov loop has the value that labels them. 

\subsection{Enforcing coincident chiral and deconfinement transitions in the phase diagram}

A challenging question is whether chiral restoration and quark deconfinement transitions coincide in the QCD phase diagram or not. In lattice QCD simulations at low chemical potentials the peak positions of chiral and Polyakov-loop susceptibilities are to good accuracy coincident, but this may not be the case for high baryon densities, allowing for three alternatives: a massive quark matter phase  \cite{Schulz:1987qg,Castorina:2008vu}, a quarkyonic phase \cite{McLerran:2007qj,Andronic:2009gj}  with chiral quarks confined in baryons (and mesons) and the case that chiral restoration and deconfinement coincide in the  
whole phase diagram.
The latter case is based on the QCD Dyson-Schwinger equations where the Kugo-Ojima confinement mechanism is realized by the dynamical quark mass function which serves for chiral symmetry breaking and at the same time conspires with the wave function renormalization so as to remove the quasiparticle pole of the quark propagator \cite{Blaschke:1997bj,Roberts:2000aa}.
Dexheimer and Schramm \cite{Dexheimer:2009va} have constructed a model that realizes coincident chiral and deconfinement phase transition via the coupling of quark and hadron masses to the Polyakov loop. The position of the critical endpoint and the critical chemical potential at $T=0$ is fixed by the behavior of the Polyakov-loop potential which they extend by adopting a chemical potential dependence. 

\begin{figure}[!t]
\includegraphics[width=0.78\linewidth]{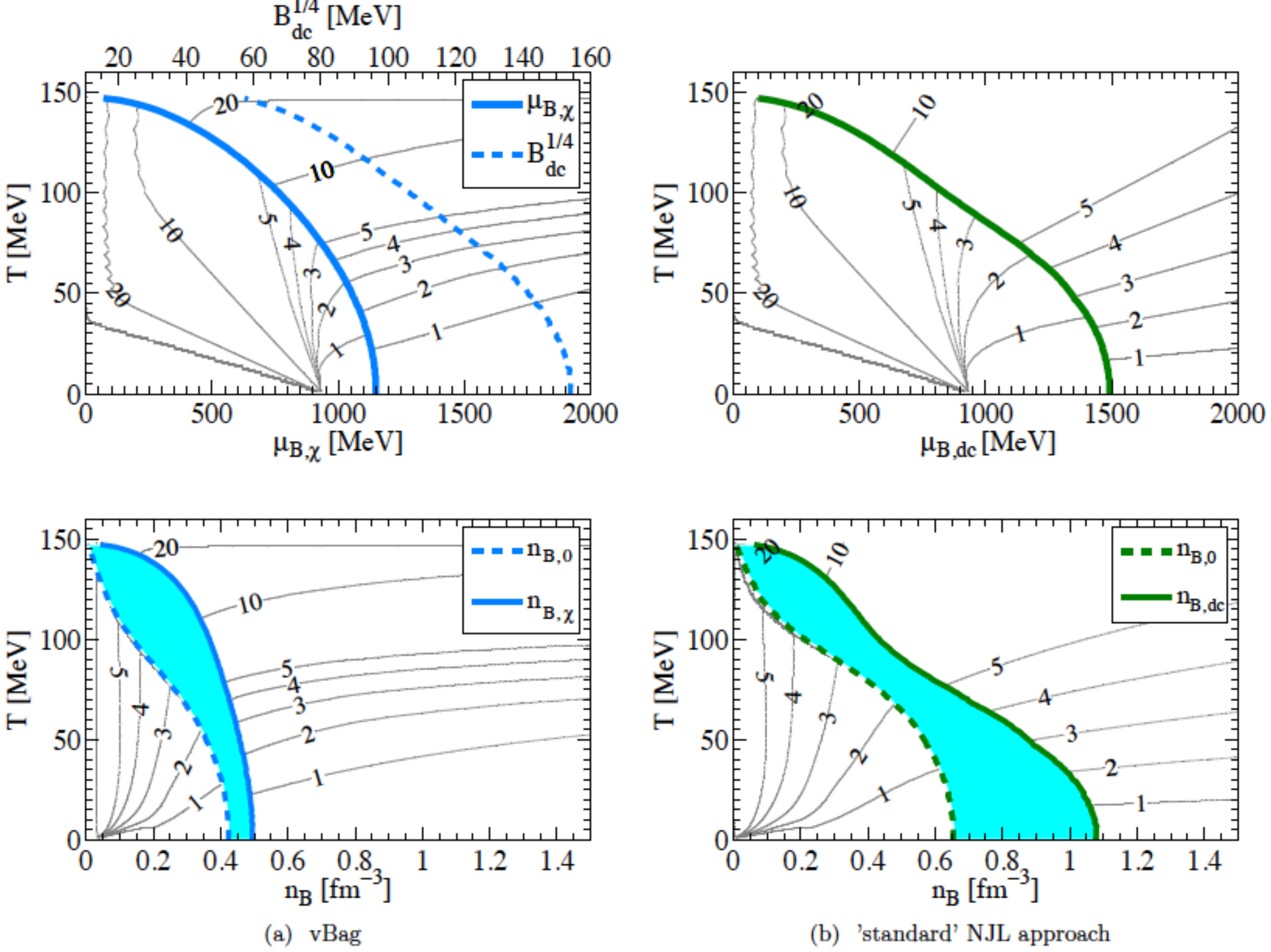}
\caption{Phase diagram for a selected chiral bag constant B$^{1/4}_\chi = 152.7$ MeV, comparing 
the new approach \cite{Klahn:2016uce,Klahn:2015mfa} (left panels) with the "standard" NJL one 
(B$_{\rm dc} = 0$, right panels). 
Top panels: Dependence on baryochemical potential. Bottom panels: Dependence on the baryon density
(see text for definitions).
Solid gray lines mark lines of constant entropy per baryon in units of the Boltzmann constant k$_{\rm B}$.
Typical trajectories of heavy-ion collisions in the NICA MPD experiment will lie between the lines labelled by "10" and "20"; for explanations see the text. Figure from Ref.~\cite{Fischer:2016ojn}.}
\label{fig-3}       
\end{figure}

In their contribution to the NICA White Paper, Fischer, Kl\"ahn and Hempel \cite{Fischer:2016ojn}
realize the coincidence of chiral restoration and deconfinement in the QCD phase diagram by adopting a bag pressure $B_{\rm dc}$, see also \cite{Klahn:2016uce}. 
The hadronic phase is described by a generalized relativistic density functional model with light clusters \cite{Typel:2009sy} and for the phase transition a Maxwell construction is applied.  In Fig.~\ref{fig-3} the phase diagram for this model (left panels) is contrasted to the case where  
$B_{\rm dc}=0$ (right panels). It is remarkable for the upcoming NICA experiments that the upper limiting density for the mixed phase is only about twice the nuclear saturation density for 
$\sqrt{s_{NN}}\sim 6$ GeV where $T\sim 130$ MeV when entering the mixed phase and after completion of the transition the matter is heated up to about 150 MeV.
For an orientation in that figure, consider the range of entropy per baryon $s/n=10 - 25$ 
\cite{Ivanov:2016hes} that will be covered by the NICA MPD collider experiment, where  
$\sqrt{s_{NN}} = 4 \dots 11$ GeV \cite{Kekelidze:2016hhw}. 
This EoS fulfils also the constraint from recent precise mass measurements of pulsars which require the maximum mass of a neutron star or hybrid star predicted for the EoS to reach at least  
$2~M_\odot$ \cite{Klahn:2015mfa}. 
\begin{figure}[!t]
\centering
\includegraphics[height=0.37\linewidth]{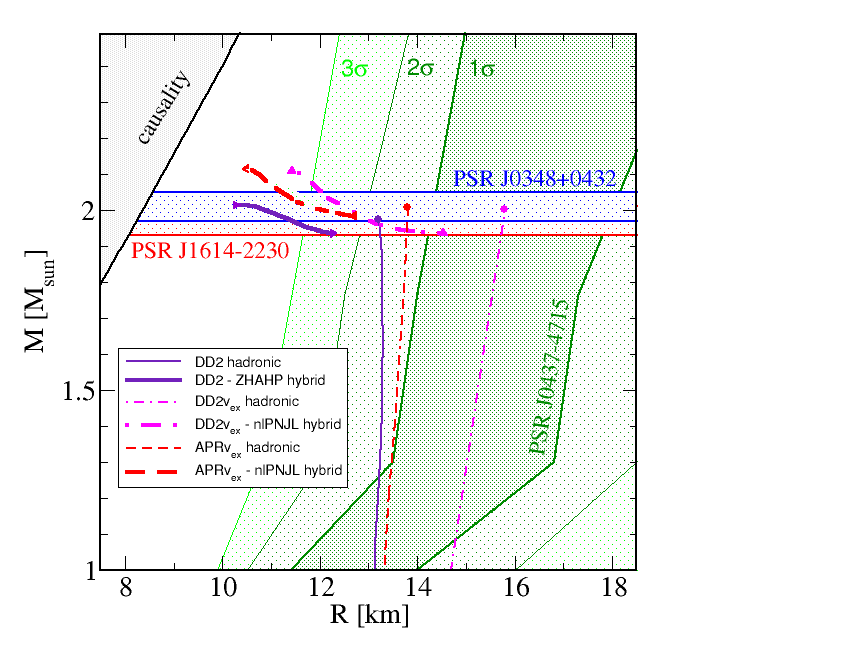}
\includegraphics[height=0.37\linewidth]{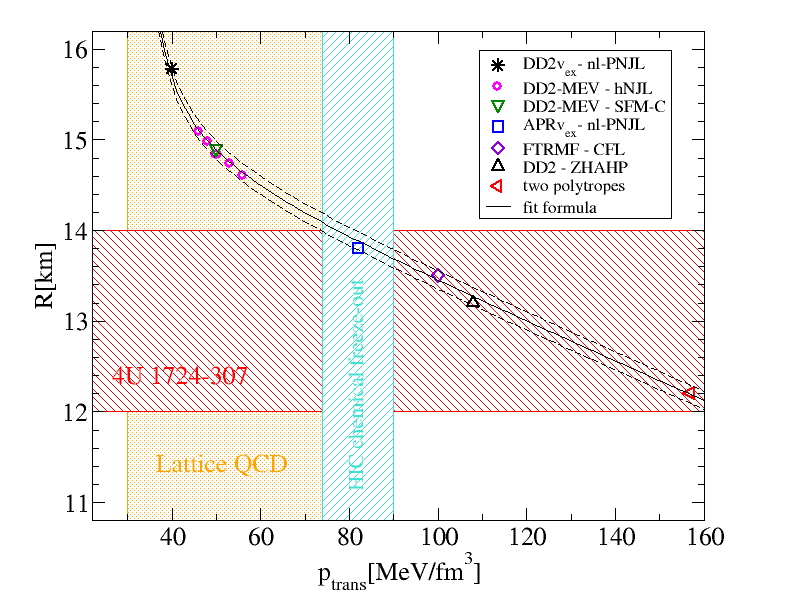}
\caption{Left panel: Mass-radius relationships for three examples of a new class of hybrid neutron star equations of state with a strong first order phase transition exhibiting an endpoint of purely hadronic configurations at about $2~M_\odot$ and a stable branch of compact hybrid stars, disconnected from the hadronic ones ("third family"). 
Right panel: The radius $R$ at the endpoint of the hadronic configurations ($\sim 2~M_\odot$) for different 
classes of hybrid equations of state versus the corresponding pressure at the phase transition 
$p_{\rm trans}$. 
All these equations of state allow for a third family of stars at high mass ("high-mass twin" stars). 
Figures from Ref.~\cite{9}.
}
\label{fig-4}       
\end{figure}

\subsection{High-mass twin stars support the CEP}

Starting from the possible existence of a third family of compact hybrid stars at high mass as indicators 
of a strong first order phase transition in compact star interiors \cite{Blaschke:2013ana} (high mass twin stars \cite{Benic:2014jia}) from which the existence of a critical point in the QCD phase diagram follows,
in the contribution \cite{9} to the NICA White Paper the relation between the pressure at the onset of the deconfinement transition for stars with $M\sim 2~M_\odot$ and their radius has been shown for the first time, see Fig.~\ref{fig-4}.      
If the maximum star radius from observations turns out to be less than 14 km, e.g. in the range 
13.6 - 14.0 km  this would indicate a limiting pressure of the hadronic phase of matter of 
74 - 90 MeV/fm$^3$, in excellent agreement with the analysis of chemical freeze-out in heavy-ion collisions 
by Rafelski \& Petran \cite{Petran:2013qla} which gave $82\pm 8$ MeV/fm$^3$. 
Such a universality of the limiting pressure of the hadronic phase in the QCD phase diagram, if it exists,  would be of enormous heuristic value for interdisciplinary studies of the QCD phase diagram in heavy-ion collisions and in Astrophysics. 
 
\section{Three-fluid hydrodynamics based simulations and mixed phase signals
\label{sec:3FH}}

\begin{figure}[!ht]
\centering
\includegraphics[width=\linewidth]{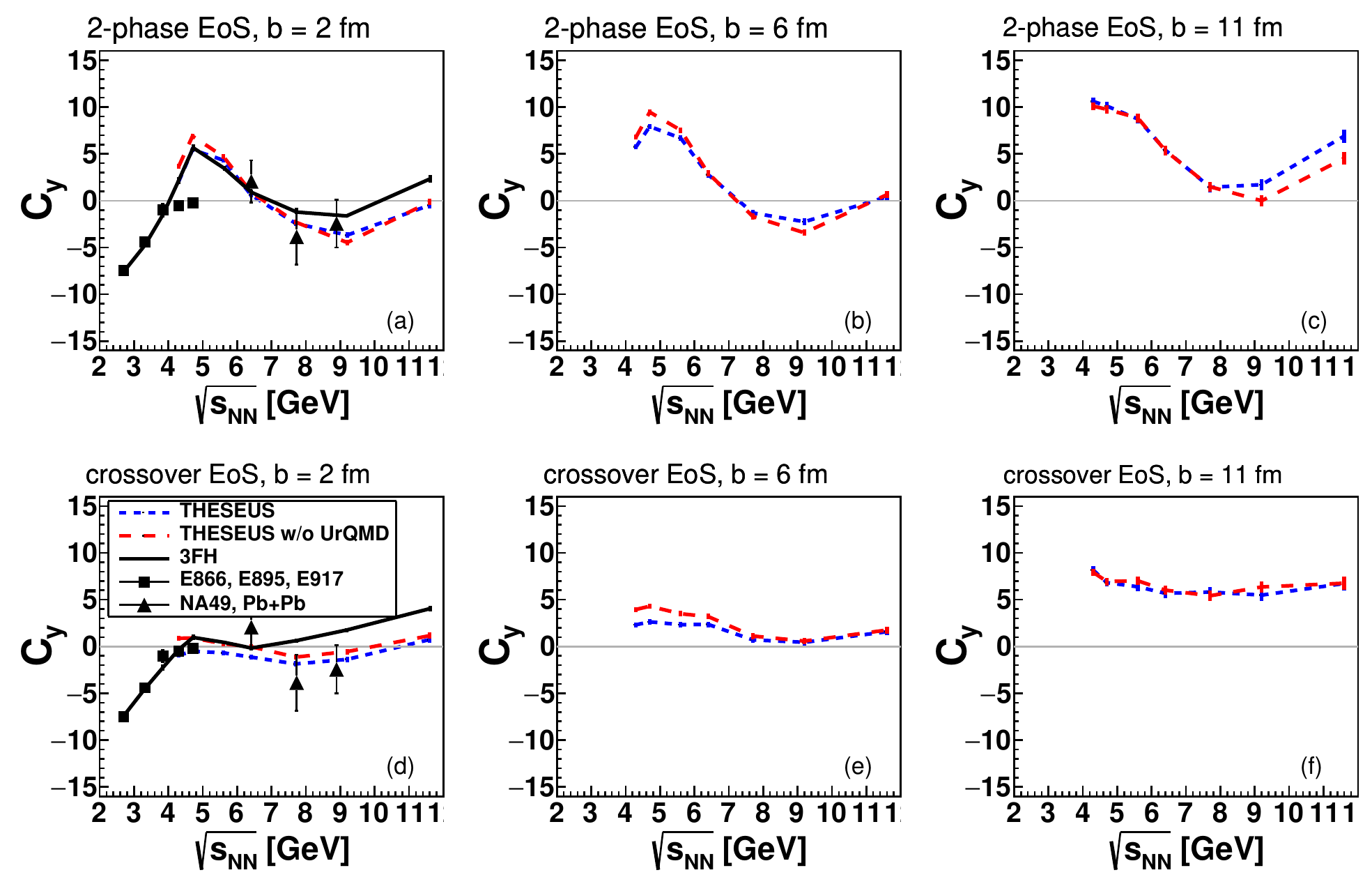}
\caption{Energy scan for the curvature Cy of the net proton rapidity distribution at midrapidity for central
Au+Au collisions with impact parameter b = 2 fm (panels (a) and (d)), b = 6 fm (panels (b) and (e)) and b = 11 fm (panels (c) and (f)). 
We compare the 3FH model result (black solid lines) with THESEUS (blue short-dashed lines) and THESEUS without UrQMD (red long-dashed lines). The results for the two-phase EoS (panels (a)-(c)) are compared to those for the crossover EoS (panels (d)-(f)). 
For noncentral collisions the curvature pattern is shifted towards positive values while the
"wiggle" as a characteristic feature for the EoS with a first order phase transition remains rather robust.
Figure from Ref.~\cite{2}.}
\label{fig-5}       
\end{figure}

\subsection{Baryon stopping signal}

In Fig.~\ref{fig-5} we show the NICA energy scan for the curvature Cy of the net proton rapidity distribution at midrapidity for central Au+Au collisions with impact parameter b = 2 fm (left panels), b = 6 fm (middle panels) and b = 11 fm (right panels). 
We compare the 3FH model result (black solid lines) with THESEUS (blue short-dashed lines) and THESEUS without UrQMD (red long-dashed lines). 
The results for the two-phase EoS (upper row of panels) are compared to those for the
crossover EoS (lower row of panels). 
For noncentral collisions the curvature pattern is shifted towards positive values while the
"wiggle" as a characteristic feature for the EoS with a first order phase transition 
\cite{Ivanov:2010cu,Ivanov:2013wha} remains rather robust \cite{Ivanov:2015vna,Ivanov:2016xev}. 
For details, see \cite{2}.

\subsection{Antiflow of light clusters}

Among the suggested signatures of a first order phase transition in heavy-ion collision experiments is the antiflow of nucleons, i.e., a negative slope for the rapidity dependence of the directed flow $v_1$ at midrapidity \cite{Stoecker:2004qu}. It has been shown in hydrodynamic calculations already in \cite{Brachmann:1999xt} that this feature is related to the softest point of the equation of state. 
Naturally, this sensitivity to the EoS could be nicely demonstrated within the three-fluid hydrodynamics approach \cite{Ivanov:2014ioa,Ivanov:2016sqy}.
The proton antiflow feature has been obtained also in different transport theory based simulations that implement a first order phase transition, for instance the hybrid approach \cite{Steinheimer:2014pfa}
which uses a hydrodynamic description of the phase transition sandwiched by UrQMD modeling of initial and final stages, and the transport theoretical model JAM \cite{Nara:2016phs,Nara:2016hbg}.

\begin{figure}[!htb]
\centering
\sidecaption
\includegraphics[height=0.67\linewidth]{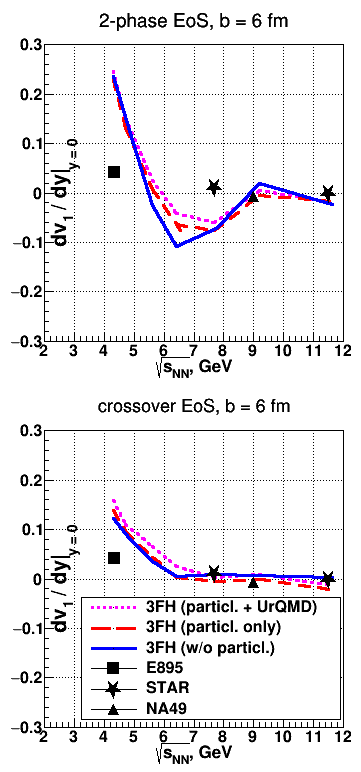}
\includegraphics[height=0.67\linewidth]{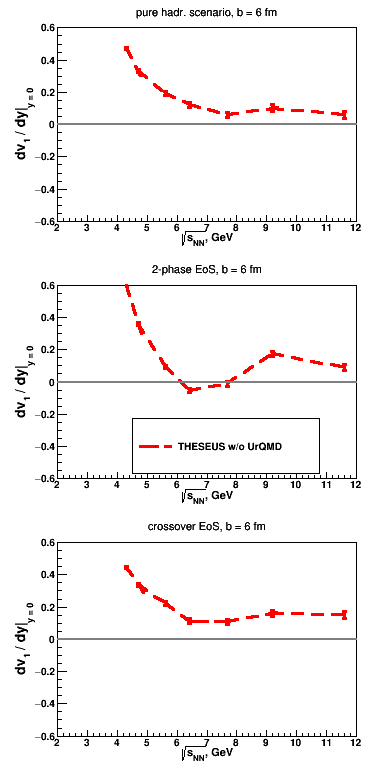}
\caption{Energy scan of the slope of the directed flow for protons (left panel) and for deuterons (right panel). A comparison of results from THESEUS simulations with a 2-phase EoS that possesses a strong first-order phase transition to results for a crossover (or purely hadronic) EoS demonstrates that a dip at energies of
$\sqrt{s_{NN}}=6-8$ GeV could be regarded as a phase transition signal. Such "antiflow" signal is expected for protons as well as light clusters. For further details, see Ref.~\cite{4}; the figures are taken from that paper.}
\label{fig-6}       
\end{figure}

In Ref.~\cite{4} it has been conjectured that the flow of light clusters such as deuterons may trace the early flow at the hadronisation transition since by their inertia the heavier particles are less affected by hadronic rescattering effects. This has been checked with the THESEUS code and indeed, the antiflow of protons 
obtained in this program for the 2-phase EoS as a signature of the first order phase transition in this EoS 
in the energy range $\sqrt{s_{NN}}\sim 6-7$ GeV is reflected also in the deuteron flow for the same EoS while the crossover and pure hadronic EoS do not show the antiflow signature, see Fig.~\ref{fig-6}.

As for the proton antiflow, it would be premature to predict quantitatively the deuteron antiflow due to present ambiguities in the deuteron formation mechanism by coalescence and in the used EoS.
In particular, the 2-phase EoS is too soft at high densities and has an unrealistically high onset density of the phase transition. With an improved EoS, the antiflow signature of the phase transition could occur in the beam energy range $4.7~{\rm GeV}<\sqrt{s_{NN}}<7.7~{\rm GeV}$, where no experimental data exist yet but where the NICA MPD experiment will be operating \cite{Kekelidze:2016hhw}.
 
\subsection{Marek's horn }

\begin{figure}[!htb]
\centering
\includegraphics[height=0.3\linewidth]{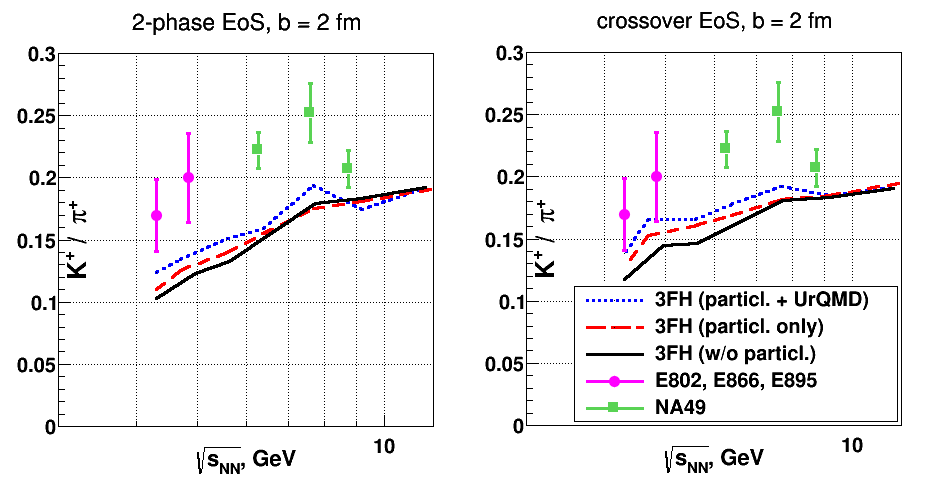}
\includegraphics[height=0.3\linewidth]{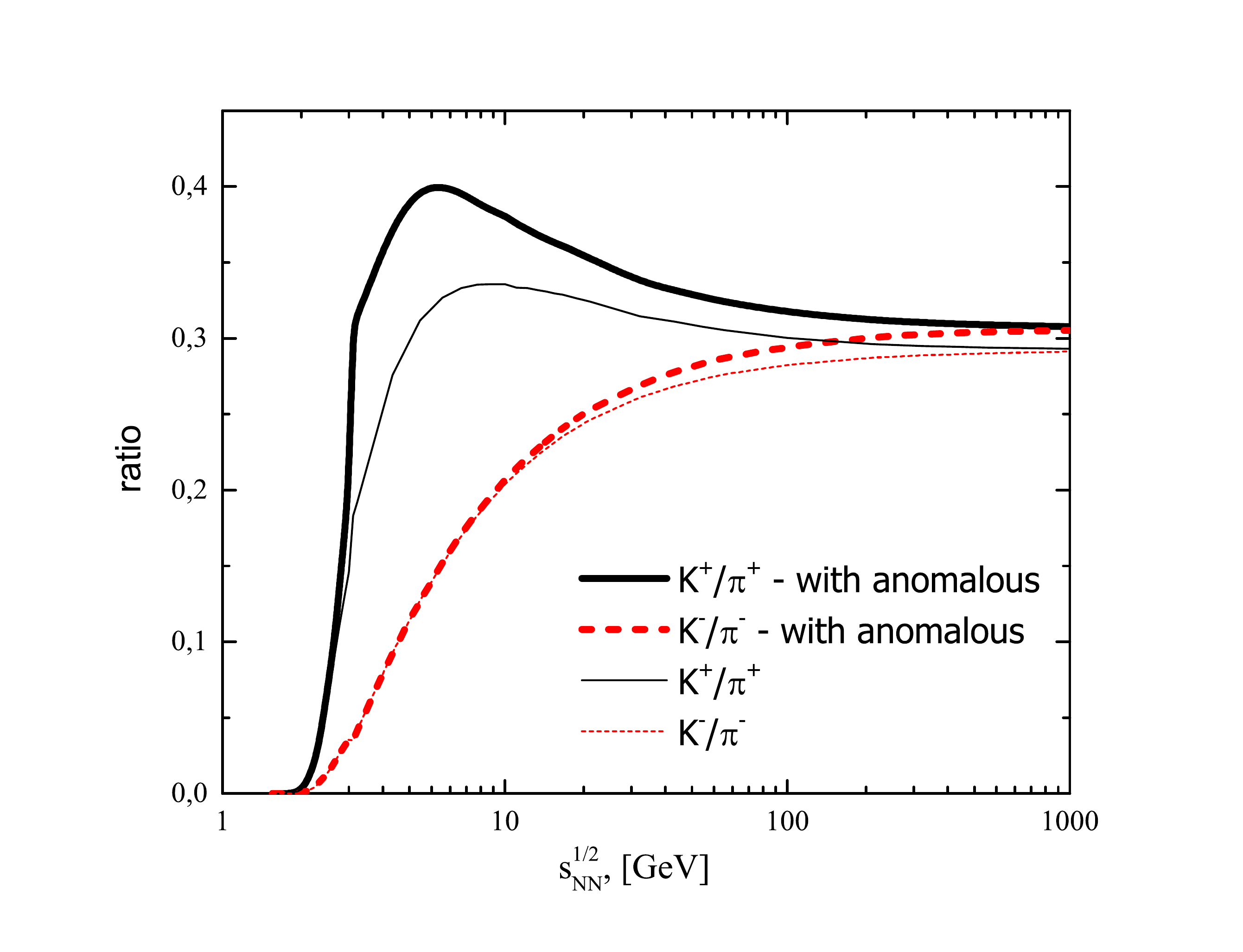}
\caption{Energy scan for the particle ratio $K^+/\pi^+$ in the NICA energy range for central Au+Au collisions (impact parameter b = 2 fm) with (blue dotted lines) and without (red dashed lines) the UrQMD hadronic rescattering. 
The calculation with a first order phase transition in the EoS (left panel) is
compared to that with the crossover EoS (middle panel). 
For comparison we show in both these panels the results without particlization and UrQMD rescattering
(black solid lines) and experimental data, taken from Fig. 11 of Ref.~\cite{Ivanov:2013yqa}.
Data from AGS experiments (E802, E866, E895) are shown by filled circles, data from NA49 by filled squares.
In the right panel we demonstrate the effect of the anomalous modes on the $K^+/\pi^+$ and $K^-/\pi^-$ ratios (thick lines), compared to the "normal" case without these modes (thin lines), from \cite{Dubinin:2016wvt}.
Upon implementation of this effect into the EoS used in the THESEUS simulation a decent improvement 
of the description of the "horn" effect will be achieved.
}
\label{fig-7}       
\end{figure}

In previous chapters we have praised the advantages of the new THESEUS simulation program for heavy-ion collisions in the NICA/FAIR energy range, where the transition from the baryon stopping to the transparency regime is expected. Promising signatures of a first order phase transition have been discussed. 
However, the "horn" effect which is observed in the energy scan of the  $K^+/\pi^+$ ratio 
(for the data see, e.g., Fig.~11 of Ref.~\cite{Ivanov:2013yqa}, partly displayed here in the left and middle panels of Fig.~\ref{fig-7}) is not reproduced in the present version of THESEUS and with the present EoS.
This effect has been suggested by Marek Gazdzicki and Mark Gorenstein as a  signature of the mixed phase in hadron-to-quark matter phase transition \cite{Gazdzicki:1998vd}.
Its quantitative description in simulations of heavy-ion collisions is a notoriously difficult problem. 
Recently, in an upgrade of the PHSD program with chiral symmetry restoration effects, a
satisfactory description of the "horn" effect could be given \cite{Palmese:2016rtq}.
We advocate another possibility to solve the "horn" puzzle.
In a recent work on the Mott dissociation of pions and kaons in hot, dense quark matter 
\cite{Dubinin:2016wvt} which extends the GBU approach \cite{Schmidt:1990,Hufner:1994ma,Blaschke:2013zaa} to the strangeness sector, we have found that a stable mode of the $K^+$ meson occurs in the phase diagram at finite $T$ and $\mu$ due to the compositeness and unequal masses of the quark constituents. 
This in-medium bound state is a good candidate for causing the enhancement of the $K^+/\pi^+$ ratio 
in a certain domain of the energy scan while the $K^-/\pi^-$ ratio remains unaffected, see the right panel 
of Fig.~\ref{fig-7}.
We plan to implement the corresponding EoS into an upgrade of the THESEUS program. 

\section{Conclusions}
In this contribution I have discussed a few selected theoretical developments of relevance for 
the NICA facility under construction.
First, I have picked new aspects of the QCD phase diagram like the possible existence of a critical endpoint of first-order phase transitions from the perspective of a nonlocal chiral quark model calibrated with lattice QCD simulations that strongly suggests a low or vanishing vector coupling at low chemical potentials; generalizations of the 3-flavor PNJL model including the conjecture of a universal pressure for the onset of deconfinement  in heavy-ion collisions and astrophysics; and a new model that enforces the coincidence of chiral and deconfinement transitions predicting a low critical density, accessible already at NICA fixed target experiments.
Next, I have discussed first results of the newly constructed event simulator THESEUS which is based on the particlization of the three-fluid hydrodynamics model by Ivanov, Russkikh and Toneev that allows also to study the role of hadronic final state interactions. Possible signals of the mixed phase accessible at NICA are considered. 
In particular, the robustness of the baryon-stopping signal of deconfinement and the occurrence of antiflow for protons and deuterons have been investigated for Au+Au collisions in the range of the NICA-MPD energy scan.
The sharp peak for the $K^+/\pi^+$ ratio at $\sqrt{s} \sim 8$ GeV (the "horn" effect) is not obtained in the present version of THESEUS and calls for improvement of the equation of state input.   
In this context, I have pointed out recent progress in developing a generalized Beth-Uhlenbeck approach to a unified description of quark-hadron matter that reveals a new mechanism for explaining the $K^+/\pi^+$ ratio. 
One may look forward to further interesting developments of theory and simulations of heavy-ion collisions in the NICA energy range directed towards the discovery of a mixed phase in the QCD phase diagram.

\subsection*{Acknowledgements}
I would like to thank the members of the board of guest editors for the EPJA Topical Issue on the "NICA White Paper", J\"org Aichelin, Elena Bratkovskaya, Volker Friese, Marek Gazdzicki, J{\o}rgen Randrup, Oleg Rogachevsky, Oleg Teryaev and Vyacheslav Toneev, for their excellent work and indispensable help in making this volume appear. 
I am grateful to all my collaborators in the research reported here, in particular, David Alvarez-Castillo, 
Alexander Ayriyan, Niels-Uwe Bastian, Pavel Batyuk, Sanjin Benic, Jens Berdermann, Marcus Bleicher, Gustavo Contrera, Pedro Costa, Pawel Danielewicz, Aleksandr Dubinin, Tobias Fischer, Hovik Grigorian, Gabriela Grunfeld, Sophia Han, Yuri Ivanov, Iuriy Karpenko,Thomas Kl\"ahn, Sergey Merts, Marlene Nahrgang, Hannah Petersen, Andrey Radzhabov, Gerd R\"opke, Oleg Rogachevsky, Ludwik Turko, 
Dmitri Voskresensky, Agnieszka Wergieluk and Hermann Wolter.  
Finally, I would like to acknowledge numerous fruitful discussions with my colleagues at JINR Dubna, especially: Valery Burov, Alexandra Friesen, Yuri Kalinovsky, Sergey Nedelko, Alexander Sorin, Oleg Teryaev, Slava Toneev and at the University of Wroclaw: Pasi Huovinen, Jakub Jankowski, Alaksiej Kachanovich, Pok Man Lo, Krzysztof Redlich, Chihiro Sasaki and Jan Sobczyk. 
This work has been supported by the MEPhI Academic Excellence Project under contract No. 02.a03.21.0005 and by the Polish NCN under grant No. UMO-2011/02/A/ST2/00306.

%
%

\end{document}